\documentclass[aps,prl,twocolumn,superscriptaddress,nofootinbib,notitlepage,longbibliography]{revtex4-1}
\usepackage{bm}
\usepackage{hhline}
\usepackage{amsmath}
\usepackage{amssymb}
\usepackage{mathdots}
\usepackage{tabularx}
\usepackage{graphicx}
\usepackage{hyperref}
\usepackage{bbding}
\usepackage{tikz}
\usepackage{bbding}
\usepackage{braket}
\usepackage{bm}

\begin{document}
 \title{Nonlinear Superconducting Magnetoelectric Effect}
	\author{Jin-Xin Hu}\thanks{jhuphy@ust.hk}
	\affiliation{Division of Physics and Applied Physics, School of Physical and Mathematical Sciences, Nanyang Technological University, Singapore 637371} 
        \author{Oles Matsyshyn}
	\affiliation{Division of Physics and Applied Physics, School of Physical and Mathematical Sciences, Nanyang Technological University, Singapore 637371}
        \author{Justin C. W. Song}\thanks{justinsong@ntu.edu.sg}
	\affiliation{Division of Physics and Applied Physics, School of Physical and Mathematical Sciences, Nanyang Technological University, Singapore 637371}

	\begin{abstract}
Supercurrent flow can induce a nonvanishing spin magnetization in noncentrosymmetric superconductors with spin-orbit interaction. Often known as the non-dissipative magnetoelectric effect, these are most commonly found at linear order in supercurrent flow. Here, we propose that a {\it nonlinear} superconducting magnetoelectric (NSM) effect can naturally manifest in magnet/superconductor heterostructures. In such platforms, NSM manifests as the spin polarization generated as a second-order response to a driving supercurrent. Strikingly, we find NSM survives centrosymmetry and is the leading order magnetic response in a variety of magnetic materials that include both collinear magnets [e.g., $d$-wave planar altermagnet thin film/superconductor (ALM/SC)] as well as noncollinear magnets (e.g., Kagome/SC systems). This renders NSM a powerful electric and non-dissipative means of controlling magnetization in magnet/superconductor heterostructures, a promising platform for superconducting spintronics.
	\end{abstract}
	\pacs{}	
	\maketitle

\emph{Introduction.}---Unlike the normal metallic state, the intrinsically non-dissipative flow of current in superconductors enables the control of spins and magnetization without dissipation~\cite{levitov1985magnetoelectric,edelstein1995magnetoelectric,edelstein2005magnetoelectric}---a critical functionality in realizing low-power spintronics~\cite{linder2015superconducting,eschrig2015spin,eschrig2011spin}. Microscopically, this process occurs through the superconducting magnetoelectric effect (sometimes known as the Edelstein effect)~\cite{edelstein1995magnetoelectric,edelstein2005magnetoelectric,fujimoto2005magnetoelectric,tkachov2017magnetoelectric,he2020magnetoelectric,he2021superconducting,ikeda2020giant,chirolli2022colossal,he2019spin,yuan2023edelstein}, where a supercurrent flowing through a noncentrosymmetric metal can induce a net magnetization. In Rashba superconductors~\cite{bauer2012non,smidman2017superconductivity,houzet2015quasiclassical}, such magnetoelectric effects can be naturally understood through the spin-orbit interaction (SOI): spin readily couples to the motion of electrons (and their associated Cooper pairs) producing a net spin polarization in the presence of supercurrent flow.

Major attention has focused on the linear superconducting magnetoelectric effect, where a spin response develops by driving supercurrent in time-reversal ($\mathcal{T}$) invariant noncentrosymmetric superconductors. For example, superconductors with a polar axis $\bm{c}$ and Rashba SOI~\cite{edelstein1995magnetoelectric,edelstein2005magnetoelectric} possess a supercurrent-induced magnetization $\bm{M}\propto \bm{c} \times \bm{J}_s$ that lies in-plane. Here $\bm{J}_s$ is the supercurrent density. However, the nonlinear magnetoelectric effect in superconductors is much less studied. Indeed, even in the normal state, nonlinear spin generation is a topic of intense recent interests~\cite{hamamoto2017nonlinear,xiao2022intrinsic,xiao2023time,guo2024extrinsic,feng2024quantum}.

\begin{figure}
		\centering
		\includegraphics[width=1.0\linewidth]{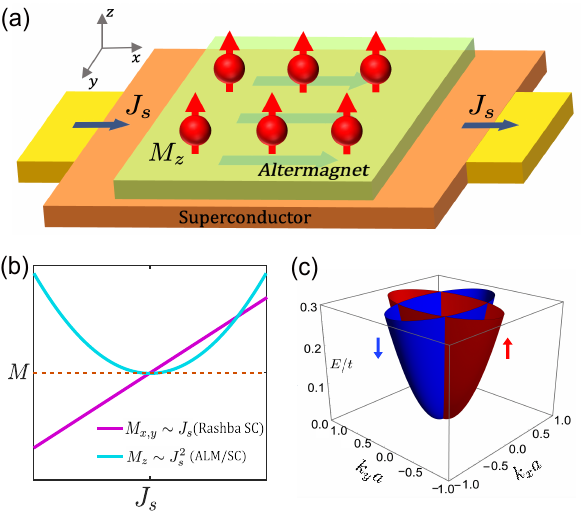}
		\caption{(a) Schematic illustration of the ALM/SC heterostructure exhibiting the nonlinear superconducting magnetoelectric (NSM) effect. In the presence of a supercurrent $J_s$ flowing along $x$ direction, the net magnetization along $z$ direction $M_z$ can be created. (b) In noncentrosymmetric superconductors with Rashba SOI, an in-plane magnetization, denoted as $M_{x,y}$, that is linear with $J_s$ can be generated. In ALM/SC heterostructures, an out-of-plane $M_z$ is generated that goes as $J_s^2$. When $J_s=0$, $M_z=0$ in the ALM/SC heterostrucure vanishes due to $C_4 \mathcal{T}$ symmetry. (c) The normal-state band structure of the altermagnet from Eq.~\eqref{eq:alm_hamiltonian} at $J_{\mathrm{ex}}=0.4t$. The spin-split bands for spin $\uparrow$ and spin $\downarrow$ are indicated in red and blue respectively.}
		\label{fig:fig1}
\end{figure}

Here we unveil the nonlinear superconducting magnetoelectric (NSM) effect by examining second-order spin generation in response to the supercurrent. Specifically, we find a (second-order nonlinear) supercurrent-induced magnetization $\delta\bm{M}$ in a two dimensional platform as 
\begin{equation}
\delta M_c^{(2)}=\chi_{ab}^c q_a q_b,
\label{eq:nonlinear_me}
\end{equation}
where $a,b=x,y$ is the direction of the supercurrent, $c=x,y,z$, and $\bm{q}$ is the Cooper pair momentum in the presence of supercurrent flow. Note $\chi_{ab}^c $ vanishes in the presence of $\mathcal{T}$ symmetry. As a result, to realize the second-order nonlinear response, $\mathcal{T}$ symmetry breaking is necessary. Interestingly, Eq.~(\ref{eq:nonlinear_me}) persists in centrosymmetric systems in sharp contrast to the case of the linear magnetoelectric effect in noncentrosymmetric superconductors~\cite{he2020magnetoelectric}. As we argue, NSM can be naturally realized in magnet/superconductor heterostructures by utilizing the proximity effect between a magnetic material and an $s$-wave superconductor~\cite{fu2008superconducting,qi2010chiral,wang2015chiral,tkachov2017magnetoelectric}. 

To be specific, we propose that NSM can dominate magnetoelectricity in a variety of magnetic material/superconductor heterostructures that include both collinear (e.g., altermagnets) as well as noncollinear (e.g., kagome lattice~\cite{lee2024fermi}) magnetic order. A particularly striking example is $d$-wave planar altermagnet thin film/superconductor (ALM/SC) heterostructure [see Fig.~\ref{fig:fig1}(a)]. Altermagnets are collinear antiferromagnets with an unconventional magnetic order~\cite{ma2021multifunctional,vsmejkal2022emerging,lee2024broken,mazin2022altermagnetism,
mazin2023altermagnetism,feng2022anomalous,vsmejkal2022giant,vsmejkal2023chiral,ghorashi2024altermagnetic,fang2024quantum,sattigeri2023altermagnetic,krempasky2024altermagnetic,fedchenko2024observation,vsmejkal2022beyond,hodt2024spin,zhu2023topological,beenakker2023phase,zhang2024finite,ouassou2023dc,sun2023andreev,papaj2023andreev,
li2023majorana,giil2024quasiclassical,sukhachov2024thermoelectric,giil2024superconductor,hodt2024interface,zyuzin2024magnetoelectric} characterized by a large spin-splitting of order hundreds of millielectronvolts~\cite{vsmejkal2022emerging}. As we describe below, the large spin-splitting in altermagnets naturally produces a pronounced second-order NSM $\chi_{ab}^z$ that dominates the out-of-plane spin magnetization of ALM/SC heterostructures. Indeed, when its $d$-wave $C_4 \mathcal{T}$ symmetry is preserved, we find $\chi_{ab}^z$ provides the {\it leading order contribution} to the out-of-plane spin magnetization sharply contrasting with polar or gyrotropic superconductors~\cite{edelstein1995magnetoelectric,he2019spin,he2020magnetoelectric} where the linear magnetoelectric effect dominates. Candidate altermagnetic materials include RuO$_2$ and KRu$_4$O$_8$ thin films that can produce sizeable NSM susceptibilities.

\emph{General theory of nonlinear superconducting magnetoelectric effect.}---We first examine the superconducting magnetoelectric effect within an effective Bogoliubov-de Gennes (BdG) framework. We note that our treatment is general and is agnostic to the precise microscopics of the superconducting states, which can be intrinsic or proximity induced. In what follows, we describe either case phenomenologically via the pairing potential $\hat{\Delta}_{\bm{k}}$. The finite-$\bm{q}$ BdG Hamiltonian in the Nambu basis $\Psi_{\bm{k},\bm{q}}=(\hat{c}_{\bm{k}+\bm{q}/2,\uparrow},\hat{c}_{\bm{k}+\bm{q}/2,\downarrow},\hat{c}^\dagger_{-\bm{k}+\bm{q}/2,\uparrow},\hat{c}^\dagger_{-\bm{k}+\bm{q}/2,\downarrow})^T$ reads
\begin{equation}
\mathcal{H}_{\mathrm{BdG}}^{\bm{k},\bm{q}}=\left(
\begin{matrix}{}
  H_{\bm{k}+\bm{q}/2} &  \hat{\Delta}_{\bm{k}}   \\
  \hat{\Delta}^\dagger_{\bm{k}}  & -H^*_{-\bm{k}+\bm{q}/2}
\end{matrix}\right),
\label{eq:bdg_alm_current}
\end{equation}
where $H_{\bm{k}}$ is the Bloch Hamiltonian in the normal state, and $\bm{q}$ is the momentum of the Cooper pair. For simplicity we have used a $\bm{q}$-independent gap function $\hat{\Delta}_{\bm{k}}$, which is valid for weak values of the applied current~\cite{chirolli2022colossal}.

The free energy $\mathcal{F}$ of the heterostructure [Fig.~\ref{fig:fig1}(a)] comprises two parts: the superconductor component $\mathcal{F}_{\mathrm{sc}}$ and the material (i.e. altermagnet) component $\mathcal{F}_{m}$ so that $\mathcal{F}=\mathcal{F}_{\mathrm{sc}}+\mathcal{F}_{m}$. The supercurrent can be tracked as $\bm{J}_s=\nu_s \bm{q}$ with the condensate superfluid stiffness $\nu_s=\partial^2 \mathcal{F}_{\mathrm{sc}}(\bm{q})/\partial^2\bm{q}$. As a result, observables that scale with $q$ below are induced by the supercurrent at linear order while $q^2$ indicates a second-order nonlinear response. 

To obtain the magnetic response, we analyze the set of parametric partition function $\mathcal{Z}(\bm{h})$
\begin{equation}
\mathcal{Z}(\bm{h})=e^{-\beta\{\mathcal{F}_{\mathrm{sc}}-\frac{1}{2\beta}\sum_{\bm{k},\bm{q},n}\mathrm{Tr}[\mathrm{log} G^{-1}(\bm{k},\bm{q},\bm{h},i\omega_n)]\}},
\label{eq:partition}
\end{equation}
where $\bm{h}$ is a parameter describing an auxiliary Zeeman field with a corresponding Zeeman energy $\Sigma(\bm{h})=\mathrm{diag}(g_s\mu_B\bm{h}\cdot\hat{s},-g_s\mu_B\bm{h}\cdot \hat{s}^*)/2$. Here $g_s=2$ is the Land\'{e} $g$ factor, $\mu_B$ is the Bohr magneton, and the Gor'kov's Green's functions are $G^{-1}(\bm{k},\bm{q},\bm{h},i\omega_n)=\mathcal{G}^{-1}(\bm{k},\bm{q},i\omega_n)+\Sigma(\bm{h})$ and $\mathcal{G}(\bm{k},\bm{q},i\omega_n)=(i\omega_n-\mathcal{H}_{\mathrm{BdG}}^{\bm{k},\bm{q}})^{-1}$ with the Matsubara frequency $\omega_n=(2n+1)\pi/\beta$. $\beta=1/k_B T$ with $T$ being the temperature. 

The spin magnetization can be evaluated as $\bm{M}=-\beta^{-1}\partial\mathrm{log}\mathcal{Z}(\bm{h})/\partial{\bm{h}}|_{\bm{h}=0}$. Using the BdG Hamiltonian [Eq.~(\ref{eq:bdg_alm_current})] in the current-carrying state, we obtain the spin magnetization as 
\begin{equation}
M_a=-\frac{g_s\mu_B}{4\beta}\sum_{n\bm{k}}\mathrm{Tr}[\mathcal{G}(\bm{k},\bm{q},i\omega_n)\eta_a].
\label{eq:bdg_magnetization}
\end{equation}
Here $\eta_a=\mathrm{diag}(s_a,-s_a^*)$ are the generalized spin Pauli matrices in the Nambu space. We then extract the linear and nonlinear responses systematically, and at linear order in $\bm{q}$ we find $\delta M_a^{(1)}=\alpha_{ab} q_b$, 
where the first-order spin susceptibility $\alpha_{ab}$ reads
\begin{equation}
\alpha_{ab}=-\frac{g_s\mu_B}{8\beta }\sum_{n\bm{k}}\mathrm{Tr}[\eta_a \mathcal{G}_0 \hat{v}_b \mathcal{G}_0], 
\label{eq:linear_sus}
\end{equation}
where $\mathcal{G}_0 \equiv \mathcal{G}(\bm{k},0,i\omega_n)=(i\omega_n-\mathcal{H}_{\mathrm{BdG}}^{\bm{k},\bm{q}=0})^{-1}$ for brevity. Here $\hat{v}_a=\mathrm{diag}[\mathcal{V}_a(\bm{k}),-\mathcal{V}_a^*(-\bm{k})]$ with $\mathcal{V}_a(\bm{k})=\partial H_{\bm{k}}/\partial k_a$. This formula is consistent with the results in Ref.~\cite{chirolli2022colossal}. Importantly, we find the nonlinear spin magnetization along $c$ direction in Eq.~(\ref{eq:nonlinear_me}) with the second-order spin susceptibility $\chi$ as
\begin{equation}
\chi_{ab}^c=-\frac{g_s\mu_B}{32\beta }\!\sum_{n\bm{k}}\!\mathrm{Tr}[\eta_c(\mathcal{G}_0\hat{w}_{ab} \mathcal{G}_0+2\mathcal{G}_0\hat{v}_{a}\mathcal{G}_0\hat{v}_{b}\mathcal{G}_0)],
\label{eq:nonlinear_sus}
\end{equation}
where $\hat{w}_{ab}=\mathrm{diag}[\mathcal{W}_{ab}(\bm{k}),-\mathcal{W}_{ab}^*(-\bm{k})]$ with $\mathcal{W}_{ab}(\bm{k})=\partial^2 H_{\bm{k}}/\partial k_a \partial k_b$. We note that $\chi_{ab}^c$ in Eq.~\eqref{eq:nonlinear_sus} is {\it general} for various model Bloch Hamiltonians and applies for both conventional pairing as well as unconventional pairing $\hat{\Delta}_{\bm{k}}$. In the following we will take a simple $s$-wave pairing $\hat{\Delta}_{\bm{k}} = \Delta_0is_y$ with a $\bm{k}$-independent pairing gap as an illustrative example to emphasize the effect of the magnetic order. More details of the derivations can be found in Supplementary Material (SM)~\cite{NoteX}.

\begin{table}
\caption{Symmetry restrictions for three components of $\bm{M}$ in a two-dimensional system. Note that $C_4\mathcal{T}$ corresponds to the case for altermagnets that we focus on in our work. Centrosymmetric $D_{3d}$ group corresponds to the kagome lattice. 1(2) indicates first (second) order response. $\checkmark$ indicates allowed and $\times$ indicates forbidden.}
\label{tab:summary}
\begin{ruledtabular}
\begin{tabular}{l|lllll}
 & $\mathcal{P}$ & $\mathcal{T}$ & $\mathcal{PT}$ & $C_4\mathcal{T}$ & $D_{3d}$ \\
 \hline
$M_x$ & \checkmark(2)      & \checkmark(1) &   $\times$ & \checkmark(1) & \checkmark(2)\\
$M_y$ & \checkmark(2)    &  \checkmark(1)   &   $\times$ & \checkmark(1) & \checkmark(2) \\
$M_z$ & \checkmark(2)     &  \checkmark(1)    &  $\times$  & \checkmark(2) & $\times$
\end{tabular}
\end{ruledtabular}
\end{table}

It is instructive to examine the symmetry requirements for both the linear $\alpha_{ab}$ and second-order NSM $\chi_{ab}^c$ summarized in Table \ref{tab:summary}. For instance, $\mathcal{T}$ breaking is necessary for non-zero $\chi_{ab}^c$ while $\mathcal{P}$ breaking is necessary for $\alpha_{ab}$; both are destroyed in $\mathcal{PT}$-invariant systems. A particular interesting case is that of $C_4\mathcal{T}$ symmetry found in altermagnets: we find that $\chi_{ab}^x$ and $\chi_{ab}^y$ terms vanish; only $\chi_{ab}^z$ is non-zero. Interestingly, $C_4\mathcal{T}$ further constrains the NSM so that $\chi_{xx}^z=-\chi_{yy}^z$. In the presence of a weak SOI naturally found in realistic altermagnets, in-plane spin magnetization can be induced as expected of a linear magnetoelectric effect in Eq.~\eqref{eq:linear_sus}. Importantly, the Rashba SOI induced linear magnetoelectric effect is purely in-plane ($\alpha_{zx}=\alpha_{zy}=0$). This means that for out-of-plane magnetic responses, the second order NSM $\chi_{ab}^z$ is the leading order term in the altermagnetic systems. As we will see below, it dominates the out-of-plane magnetic responses. We note that Ref.~\cite{zyuzin2024magnetoelectric} recently discussed how $d$-wave magnetic order in a superconductor induces a second-order NSM response. Our general formulation reveals that such NSM responses also extend beyond altermagnets. An illustrative example involves materials in the $D_{3d}$ group; in such systems, we find that NSM persists and dominates the in-plane magnetic responses, see below.

\begin{figure}
		\centering
		\includegraphics[width=1.0\linewidth]{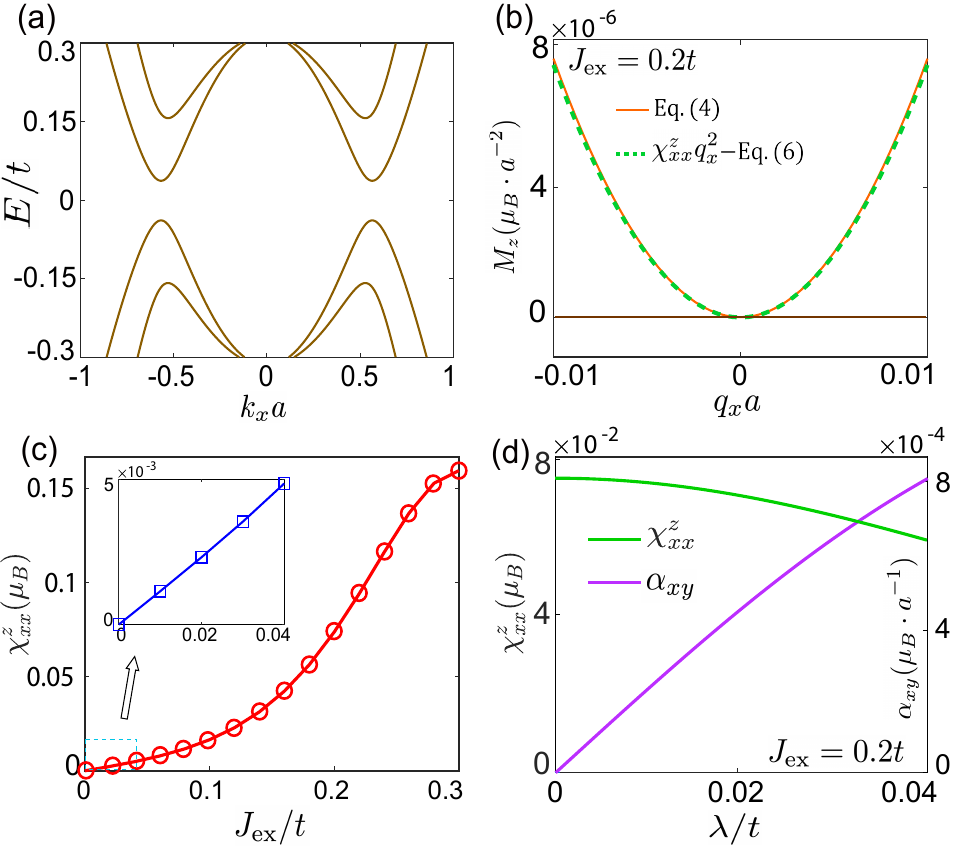}
		\caption{(a) The BdG energy spectrum of ALM/SC from Eq.~\eqref{eq:bdg_alm_current} with Bloch Hamilotnian in Eq.~\eqref{eq:alm_hamiltonian} at $k_y=0$ with spin-singlet pairing $\Delta_0$. (b) The spin magnetization $M_z$ as a function of $q_x$ from Eq.~\eqref{eq:bdg_magnetization} (solid orange). As a comparison, the NSM susceptibility multiplied by $q_x^2$ is plotted as Eq.~\eqref{eq:nonlinear_sus} (dashed green). (c) The nonlinear spin susceptibility $\chi^z_{xx}$ as a function of $d$-wave magnetic order $J_{\mathrm{ex}}$. The inset shows the region of $0<J_{\mathrm{ex}}<0.04t$. (d) For nonzero $\lambda$, the linear and nonlinear spin susceptibilities $\alpha_{xy}$ and $\chi^z_{xx}$ as a function of $\lambda$. $\alpha_{xy}$ increases linearly with $\lambda$ for small $\lambda$.  Parameters: $\lambda=0$ for (a), (b) and (c). $J_{\mathrm{ex}}=0.2t$ for (a), (b) and (d). $(\Delta_0,\mu)=(0.1,0.3)t$ for all four panels. The temperature is set to be $T=0.3T_c$. In our calculations, we adopt the BCS temperature dependence of $\Delta_0$ with $\Delta(T)=\Delta_0\tanh(1.74\sqrt{T_c/T-1})$ and $\Delta_0=1.76k_B T_c$. }
		\label{fig:fig2}
\end{figure}
\emph{NSM in ALM/SC heterostructure.}---To exemplify NSM, we first focus on the ALM/SC set-up illustrated in Fig.~\ref{fig:fig1}(a): an altermagnet thin film is in contact with a conventional $s$-wave superconductor. To characterize the physics of quasi-2D planar $d_{x^2-y^2}$-wave altermagnet, we adopt a generic two-band Hamiltonian~\cite{vsmejkal2022emerging}, yielding
\begin{equation}
H_{\bm{k}}=(ta^2\bm{k}^2-\mu)s_0+J_{\mathrm{ex}}a^2( k_x^2 - k_y^2)s_z.
\label{eq:alm_hamiltonian}
\end{equation}
Here $t$ parameterizes the usual kinetic energy, $J_{\mathrm{ex}}$ denotes the $d_{x^2-y^2}$ exchange magnetic order and $\mu$ is the Fermi energy. $a$ denotes the lattice constant. We note that an additional Rashba SOI term $H_R=\lambda a(k_x s_y-k_y s_x)$ can be readily added to $H_{\bm{k}}$ with $\lambda$ term representing the strength of SOI, that arises in altermagnets~\cite{vsmejkal2022emerging,fang2024quantum}. This Hamiltonian breaks time reversal symmetry $\mathcal{T}=-i s_y\mathcal{K}$ ($\mathcal{K}$ is the complex conjugate) and four-fold rotation symmetry $C_4=e^{i\pi s_z/4}$, but preserves the combination of them as $C_4 \mathcal{T}$. The SOI term breaks inversion ($\mathcal{P}$) symmetry without breaking $C_4 \mathcal{T}$. Throughout our work, we use the energy unit $t$. As an illustration of the spin-split band structure in altermagnet, we plot Eq.~(\ref{eq:alm_hamiltonian}) in Fig.~\ref{fig:fig1}(c); here we have used $J_{\mathrm{ex}}=0.4t$ and $\lambda=0$.
  
Because of the proximity effect between the altermagnet and superconductor, Cooper pairs can tunnel into the altermagnet. For a thin film of altermagnet [see Fig.~\ref{fig:fig1}(a)], with thickness $d$ much less than the coherence length of the superconductor ($d \ll \xi$), the pairing gap is approximately uniform along the $z$ direction. We have focused on an conventional $s$-wave superconductor with the pairing $\Delta_0$. Using the parameters as $(J_{\mathrm{ex}},\lambda,\mu,\Delta_0)=(0.2,0,0.3,0.1)t$, we diagonalize the BdG Hamiltonian and plot the fully gapped BdG energy spectrum in Fig.~\ref{fig:fig2}(a). Although the pairing potential is purely $s$-wave, the induced superconducting correlations can be obtained as $F(\bm{k},\omega)=[\psi_0+i\omega\bm{d}(\bm{k})\cdot \bm{s}]is_y$, where $\psi_0$ and $\bm{d}$ vector parameterize the spin-singlet and spin-triplet pairing correlations, respectively~\cite{bergeret2005odd,gor2001superconducting,frigeri2004superconductivity,zhou2016ising,linder2019odd}. We have $d = (0,0,d_z)$ with $d_z \propto J_{\mathrm{ex}}(k_x^2-k_y^2)$, indicating that the $d$-wave magnetic order gives rise to the odd-frequency spin-triplet pairing correlation in altermagnets due to the proximity effect.

To illustrate NSM effect in the ALM/SC heterostructure, we numerically study the supercurrent induced magnetization in Fig.~\ref{fig:fig2} for the Bloch Hamiltonian in Eq.~\eqref{eq:alm_hamiltonian}. By applying the current along $x$ direction as $\bm{q}=(q_x,0)$, we plot the magnetization directly from Eq.~\eqref{eq:bdg_magnetization} at $J_{\mathrm{ex}}=0.2t$ in Fig.~\ref{fig:fig2}(b). We find that $M_z\sim q_x^2$ as expected from Fig.~\ref{fig:fig1}(b). It is also worth noting that when $q_x=0$, equilibrium magnetization vanishes  $M_z=0$, which is consistent with $C_4 \mathcal{T}$ symmetry in altermagnets. We note that numerically computing the NSM susceptibility $\chi_{xx}^z$ from Eq.~\eqref{eq:nonlinear_sus} and multiplying by $q_x^2$ as plotted in the dashed green line matches the solid line in orange [from Eq.~\eqref{eq:bdg_magnetization}]. 

In Fig.~\ref{fig:fig2}(c), we show the second-order spin susceptibility $\chi_{xx}^z$ as a function of $J_{\mathrm{ex}}$ from Eq.~\eqref{eq:nonlinear_sus}. Generally, $\chi_{xx}^z$ starts to grow linearly with $J_{\mathrm{ex}}$ as shown in the zoom-in inset of Fig.~\ref{fig:fig2}(c). The slope ($\partial\chi_{xx}^z/\partial J_{\mathrm{ex}}$) can be derived analytically as $\chi_{xx}^z=\mu_B J_{\mathrm{ex}}\int_0^\infty f(\zeta)d\zeta + \mathcal{O}(J_{\rm ex}^2)$ with 
\begin{equation}
\begin{split}
f(\zeta)=&\frac{\beta^3}{64\pi} \zeta\mathrm{sech}^2\gamma \Big[\frac{4}{\beta^2}+t^2\zeta^4(2-3\mathrm{sech}^2\gamma)\\
&+\frac{4t\zeta^2(\mu-t\zeta^2)\tanh\gamma}{\gamma}\Big],
\end{split}
\label{eq:sus_tem}
\end{equation}
where $\gamma=\beta\sqrt{\Delta(T)^2+(t\zeta^2-\mu)^2}/2$. Furthermore, we also calculate the first-order spin susceptibility $\alpha_{xy}$ (purple) which is shown in Fig.~\ref{fig:fig2}(d). As expected, $\alpha_{xy}$ directly depends on $\lambda$; for small $\lambda$, $\alpha_{xy}$ increases linearly with $\lambda$ and vanishes when $\lambda=0$. 

The role of $\lambda$ (and more generally other SOI) for NSM $\chi_{xx}^z$ is more nuanced. At zero temperature, $\chi_{xx}^z$ vanishes as $\chi_{xx}^z\sim \beta^3\mathrm{sech}^2\beta$ [Eq.~\eqref{eq:sus_tem}] when $\lambda =0$ since $[\eta_z,\mathcal{H}_{\mathrm{BdG}}^{\bm{k},\bm{q}}]=0$ rendering $\langle \eta_z\rangle $ conserved. When $\lambda\neq 0$ is turned on, $\langle \eta_z\rangle $ is no longer conserved turning on $\chi_{xx}^z$. In contrast, at finite temperature $\chi_{xx}^z$ is sizeable even for $\lambda=0$ yielding a nonzero $M_z$ that arises from thermal fluctuations at finite temperature in the current-carrying state. Indeed, $\chi_{xx}^z$ at finite temperature (e.g.,  $T=0.3 T_{\rm c}$) in Fig.~\ref{fig:fig2}(d) exhibits a weak dependence on $\lambda$. This evidences the large altermagentic spin splitting~\cite{vsmejkal2022emerging} of order several hundred millielectronvolts captured by $J_{\rm ex} \gg \lambda$ dominates the magnetic order yielding a robust NSM.

\emph{NSM in a noncollinear antiferromagnet.}---So far we have focused on NSM in $d$-wave altermagnets---a type of collinear antiferromagnet. The general NSM in Eq.~\eqref{eq:nonlinear_sus} can exist in a larger class of magnet/superconductor heterostructures (see e.g., Table \ref{tab:summary} as well as SM~\cite{NoteX} for a detailed discussion). Indeed, based on Table \ref{tab:summary}, we find that for centrosymmetric noncollinear antiferromagnet with $D_{3d}$ symmetry, an in-plane NSM with $\chi_{xx}^y=\chi_{xy}^x=\chi_{yx}^x=-\chi_{yy}^y\neq 0$ can manifest. To capture the physics of a metal with noncollinear magnetic order on a kagome lattice, we adopt the low energy Hamiltonian~\cite{lee2024fermi,NoteX} as
\begin{equation}
\label{eq:d3d_hamil}
H_{\bm{k}}=(ta^2\bm{k}^2-\mu)s_0+m a^2[2k_x k_y s_x+(k_x^2 - k_y^2)s_y], 
\end{equation}
where $m$ captures the noncollinear magnetic order. Note that the noncollinear magnetic order induced winding spin texture on the Fermi surface [Fig.~\ref{fig:fig3}(a)] sharply contrasts with the collinear magnetic order in $d$-wave altermagnets [Fig.~\ref{fig:fig1}(c)]. As we now explain, this difference in magnetic order leads to an in-plane NSM.

To see this, we follow the same procedure as described above and utilize the parameters $(m,\mu,\Delta_0)=(0.2,0.3,0.1)t$. We obtain an in-plane NSM in Fig.~\ref{fig:fig3}(b) that dominates the magnetoelectricity; indeed, as expected from Table.~\ref{tab:summary} (last column), $M_y$ is quadratic in $q_x$ -- it is the leading order in-plane magnetoelectric term. In-plane NSM is directly controlled by the noncollinear magnetic order as evidenced in Fig.~\ref{fig:fig3}(c), where $\chi_{xx}^y$ turns on with $m$. As a result, we find that NSM arises in the presence of a variety of magnetic orders: such as $J_{\mathrm{ex}}$ (altermagnet, collinear) or $m$ (kagome, noncollinear) yielding both out-of-plane as well as in-plane NSM responses.

\begin{figure}
		\centering
		\includegraphics[width=1\linewidth]{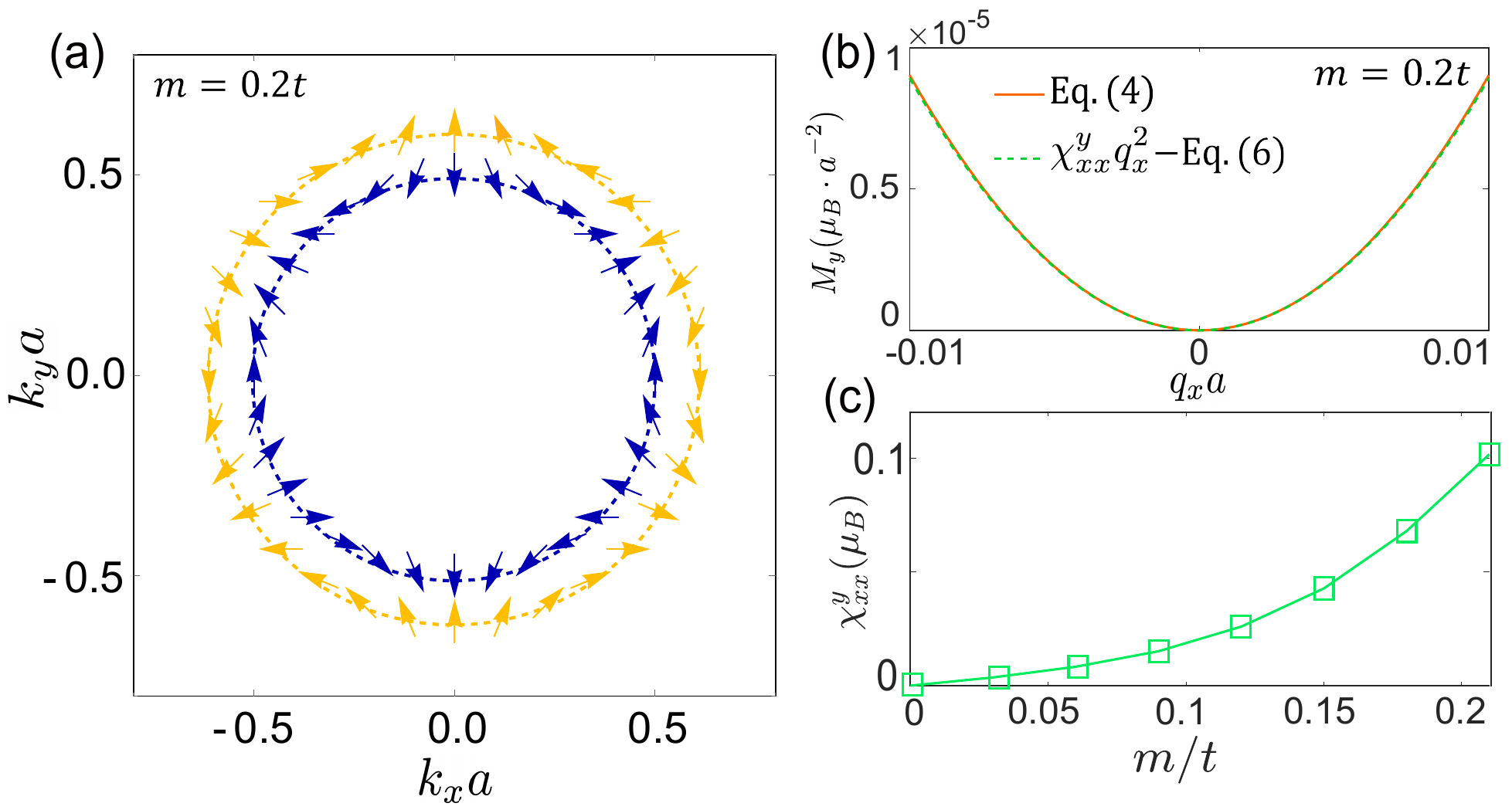}
		\caption{(a) The Fermi surface associated with the winding spin texture of Eq.~\eqref{eq:d3d_hamil}. Blue (yellow) color denotes the upper (lower) band. (b) The spin magnetization $M_y$ as a function of $q_x$ from Eq.~\eqref{eq:bdg_magnetization} (solid orange) aligns with $\chi^y_{xx}q_x^2$ from Eq.~\eqref{eq:nonlinear_sus} (dashed green). (c) $\chi^y_{xx}$ as a function of noncollinear magnetic order $m$. We set $(\Delta_0,\mu)=(0.1,0.3)t$ and $T=0.3T_c$.}
		\label{fig:fig3}
\end{figure}
\emph{Discussion.}---We expect that NSM can be realized in several concrete materials. As an example, considering the altermagnet RuO$_2$ thin film with lattice constant $a=4.5\AA$, $c=3.1\AA$, and $t=2.5$ eV, $J_{\mathrm{ex}}=0.5$ eV~\cite{vsmejkal2022giant,zhang2024finite}. Using a pairing gap of $\Delta_0=1$ meV, $\mu=0.1$ eV, $\lambda=40$ meV and $T=0.3T_c$ we estimate $\chi_{xx}^z\approx 0.1 \mu_B$. For a Cooper pair momentum $q\approx  10^{-3}\AA^{-1}$~\cite{zhu2021discovery}, the induced out-of-plane spin density in the RuO$_2$ thin film/superconductor heterostructure is $\sim 0.3\times 10^{-4}\mu_B/\mathrm{nm}^3$. Note that this value is comparable to the spin generation in MnBi$_2$Te$_4$ and other noncentrosymmetric ferromagnetic systems~\cite{xiao2022intrinsic,chernyshov2009evidence,fang2011spin}. Another example is KRu$_4$O$_8$ with the parameter values: $t=0.05$ eV, $J_{\mathrm{ex}}=0.018$ eV, and $a=9.9\AA$~\cite{zhang2024finite}. We find $\chi_{xx}^z\approx 0.01\mu_B$ and out-of-plane spin density similar to that discussed above. The induced NSM magnetization densities (of order $\sim 10^{-4}\mu_B/\mathrm{nm}^3$ corresponding to equivalent magnetic fields $\sim 1 \mu {\rm T}$) can be readily detected by superconducting quantum interference devices (SQUID)~\cite{chirolli2022colossal,vasyukov2013scanning} or diamond nitrogen-vacancy center-based magnetometers~\cite{balasubramanian2009ultralong}. 

Naturally, high quality interfaces should be used, since disorder can locally disrupt the superconducting pairing and suppress the supercurrent. Additionally, in our analysis above, we have focussed on ideal ALM/SC heterostructures; in realistic systems, surface reconstruction may ensue. For example, reconstruction can produce to interfacial spin-orbit coupling that breaks local mirror symmetry~\cite{linder2011spin}, inducing additional linear magnetoelectric responses including in the out-of-plane direction (see SM~\cite{NoteX} of interface effects). Nevertheless, we expect the large altermagnetic spin-splitting~\cite{vsmejkal2022emerging} readily dominates the response rendering NSM robust.

Even in the presence of an out-of-plane linear magnetoelectric response, we anticipate that NSM can be distinguished by {\it dynamical probes}: e.g., oscillatory Cooper pair momentum $\bm{q} (t) =\bm{q} \cos\omega t$ for slowly varying AC supercurrent with $\omega\ll\Delta_{0}$. In this weak oscillating regime, a rectified as well as a second harmonic magnetization appear: $\delta M_c^{(2)} (t) = \chi^c_{ab} q_a q_b (1 + \cos2\omega t)/2 $. Both rectified as well as second harmonic depart from that of the conventional linear magnetoelectric effect, which produces an oscillating magnetization at frequency $\omega$. This distinction is important practically since it enables to directly identify NSM even in systems where NSM is not the leading order magnetoelectric effect: e.g.,  conventional linear magnetoelectric effect produces a magnetization that averages to zero over one cycle, while the rectified NSM does not vanish allowing to directly identify its contribution; similarly locking-in to $2\omega$ allows to directly distinguish between linear and nonlinear superconducting magnetoelectric effects.

While NSM is naturally realized in ALM/SC heterostructures  (driven by its large spin-splitting), we find that it persists more generally for a wide range of other materials and symmetries (see SM~\cite{NoteX}). Strikingly, when NSM is combined with the linear magnetoelectric effect, current induced magnetization can become {\it non-reciprocal} producing distinct magnetizations for $+\bm{q}$ and $-\bm{q}$~\cite{NoteX}, a new tool for non-dissipative spintronics. Perhaps most interesting, however, is the NSM's ability to distinguish dilineate magnetic orders: as a result, we expect NSM provides a promising approach for exploring the rich interplay between magnetism and superconductivity in unconventional magnetic materials that include kagome materials~\cite{yin2022topological}, as well as materials in different spin space groups~\cite{xiao2024spin,jiang2024enumeration,chen2024enumeration}.

\emph{Acknowledgements.}---We thank Bo Yang, Wen-Yu He, Naizhou Wang and Benjamin T. Zhou for helpful discussions. This work was supported by the Ministry of Education Singapore under its Academic Research Fund Tier 2 Grant No. MOE-T2EP50222-0011.

\clearpage
		\onecolumngrid
\begin{center}
		\textbf{\large Supplementary Material for\\ ``Nonlinear Superconducting Magnetoelectric Effect''}\\[.2cm]		
      Jin-Xin Hu,$^{1}$  Oles Matsyshyn,$^{1}$   Justin C. W. Song$^{1}$\\[.1cm]
        {\itshape ${}^1$Division of Physics and Applied Physics, School of Physical and Mathematical Sciences, Nanyang Technological University, Singapore 637371}

\end{center}
	
	\maketitle

\setcounter{equation}{0}
\setcounter{section}{0}
\setcounter{figure}{0}
\setcounter{table}{0}
\setcounter{page}{1}
\renewcommand{\theequation}{S\arabic{equation}}
\renewcommand{\thesection}{ \Roman{section}}

\renewcommand{\thefigure}{S\arabic{figure}}
\renewcommand{\thetable}{\arabic{table}}
\renewcommand{\tablename}{Supplementary Table}

\renewcommand{\bibnumfmt}[1]{[S#1]}
\renewcommand{\citenumfont}[1]{#1}
\makeatletter

\maketitle

\setcounter{equation}{0}
\setcounter{section}{0}
\setcounter{figure}{0}
\setcounter{table}{0}
\setcounter{page}{1}
\renewcommand{\theequation}{S\arabic{equation}}
\renewcommand{\thesection}{ \Roman{section}}

\renewcommand{\thefigure}{S\arabic{figure}}
\renewcommand{\thetable}{\arabic{table}}
\renewcommand{\tablename}{Supplementary Table}

\renewcommand{\bibnumfmt}[1]{[S#1]}
\renewcommand{\thesection}{S\arabic{section}}
\renewcommand{\theequation}{S\arabic{equation}}
\renewcommand{\thetable}{S\arabic{table}}
\renewcommand{\thefigure}{S\arabic{figure}}
\setcounter{equation}{0}
\setcounter{page}{1}

\maketitle

\makeatletter 

\section{Microscopic theory of superconducting magnetoelectric effect}
\subsection{Green's function method}
We provide more details in deriving the general formula of superconducting magnetoelectric effect, including both the linear and nonlinear terms. The superconducting state can be intrinsic or proximity induced with a pairing function $\hat{\Delta}_{\bm{k}}$. $\hat{\Delta}_{\bm{k}}$ can be conventional $s$-wave with a uniform pairing gap $\hat{\Delta}_{\bm{k}}=\Delta_0is_y$ or unconventional pairing states. In the current-carrying state, the finite-$\bm{q}$ mean field Hamiltonian is $H_{\mathrm{mf}}(\bm{q})=\frac{1}{2}\Psi_{\bm{k},\bm{q}}^\dagger \mathcal{H}_{\mathrm{BdG}}^{\bm{k},\bm{q}}\Psi_{\bm{k},\bm{q}}$ in the Nambu basis $\Psi_{\bm{k},\bm{q}}=(\hat{c}_{\bm{k}+\bm{q}/2,\uparrow},\hat{c}_{\bm{k}+\bm{q}/2,\downarrow},\hat{c}^\dagger_{-\bm{k}+\bm{q}/2,\uparrow},\hat{c}^\dagger_{-\bm{k}+\bm{q}/2,\downarrow})^T$. The Bogoliubov-de Gennes (BdG) Hamiltonian reads
\begin{equation}
\mathcal{H}_{\mathrm{BdG}}^{\bm{k},\bm{q}}=\left(
\begin{matrix}{}
  H_{\bm{k}+\bm{q}/2} &   \hat{\Delta}_{\bm{k}}  \\
  \hat{\Delta}^\dagger_{\bm{k}}  & -H^*_{-\bm{k}+\bm{q}/2}
\end{matrix}\right),
\label{eq:bdg_alm_current_supple}
\end{equation}
where $H_{\bm{k}}$ is the Bloch Hamiltonian of the normal state, $\bm{q}$ is the momentum of the Cooper pair, and $s_y$ is a Pauli matrix. For simplicity we have used a $\bm{q}$-independent gap function $\hat{\Delta}_{\bm{k}}$, which is valid for weak values of the applied current.

In the main text, we focus on the altermagnet/superconductor heterostructure. Therefore the total free energy $\mathcal{F}$ of the system contains the superconductor part $\mathcal{F}_{\mathrm{sc}}$ and the material part $\mathcal{F}_{m}$ with $\mathcal{F}=\mathcal{F}_{\mathrm{sc}}+\mathcal{F}_{\mathrm{m}}$. The supercurrent $J_s=\nu_s \bm{q}$ with the superfluid density $\nu_s=\partial^2 \mathcal{F}_{\mathrm{sc}}(\bm{q})/\partial^2\bm{q}$. The partition function $\mathcal{Z}(\bm{h})$ is given by

\begin{equation}
\mathcal{Z}(\bm{h})=e^{-\beta[\mathcal{F}_{\mathrm{sc}}-\frac{1}{2\beta}\sum_{\bm{k},\bm{q},n}\mathrm{Tr}[\mathrm{log} G^{-1}(\bm{k},\bm{q},\bm{h},i\omega_n)]]}.
\end{equation}
By introducing an auxiliary Zeeman field $\bm{h}$, we can write down the modified BdG Hamiltonian:
\begin{equation}
\mathcal{H}_{\mathrm{BdG}}^{\bm{k},\bm{q}}(\bm{h})=
\left(\begin{array}{cc}
  H_{\bm{k}+\bm{q}/2}-\frac{1}{2}g_s \mu_B \bm{h}\cdot \bm{s} & \hat{\Delta}_{\bm{k}} \\
\hat{\Delta}^\dagger_{\bm{k}} & -H^*_{-\bm{k}+\bm{q}/2}+\frac{1}{2}g_s \mu_B \bm{h}\cdot \bm{s}^*
\end{array}\right),
\end{equation}
This auxiliary Zeeman field is to be distinguished from the
genuine magnetization of the system and is set to zero at
the end of the calculation. The Gor'kov's Green's function is $G(\bm{k},\bm{q},\bm{h},i\omega_n)=[i\omega_n-\mathcal{H}_{\mathrm{BdG}}^{\bm{k},\bm{q}}(\bm{h})]^{-1}$. The magnetization $M$ can be obtained from the partition function as
\begin{equation}
\begin{split}
M_a&=-\frac{1}{\beta}\frac{\partial}{\partial {h_a}}\mathrm{log}Z(\bm{h})\\
&=-\frac{1}{2\beta}\sum_{\bm{k},\bm{q},n}\frac{\partial}{\partial {h_a}}\mathrm{Tr}[\mathrm{log}G^{-1}(\bm{k},\bm{q},\bm{h},i\omega_n)].
\end{split}
\end{equation}
We decompose the Green’s function as $G^{-1}(\bm{k},\bm{q},\bm{h},i\omega_n)=\mathcal{G}^{-1}(\bm{k},\bm{q},i\omega_n)+\Sigma(\bm{h})$ with 
\begin{equation}
\Sigma(\bm{h})=\left(\begin{array}{cc}
 \frac{1}{2}g_s \mu_B \bm{h}\cdot \bm{s} & 0 \\
0 & -\frac{1}{2}g_s \mu_B \bm{h}\cdot \bm{s}
\end{array}\right).
\end{equation}
We can then expand the logarithm in the first order of $\bm{h}$ as $\mathrm{log}G^{-1}(\bm{k},\bm{q},\bm{h},i\omega_n)=\mathrm{log}\mathcal{G}^{-1}(\bm{k},\bm{q},i\omega_n)+\mathcal{G}(\bm{k},\bm{q},i\omega_n)\Sigma(\bm{h})$. Thus, we can then obtain the magnetization $M$ as
\begin{equation}
\label{eq:magnetization}
M_a=-\frac{g_s\mu_B}{4\beta}\sum_{n\bm{k}}\mathrm{Tr}[\mathcal{G}(\bm{k},\bm{q},i\omega_n)\eta_a].
\end{equation}
Here $\eta_a=\mathrm{diag}(s_a,-s_a^*)$ is the redefined spin Pauli matrices in the Nambu space. The Gor'kov Green's function is $\mathcal{G}(\bm{k},\bm{q},i\omega_n)=(i\omega_n-\mathcal{H}_{\mathrm{BdG}}^{\bm{k},\bm{q}})^{-1}$ with the Matsubara frequency $\omega_n=(2n+1)\pi k_B T$. $T$ is the temperature. $g_s=2$ is the Land\'{e} $g$ factor and $\mu_B$ is the Bohr magneton. We can then expand the BdG Hamiltonian as 
\begin{equation}
\label{eq:HBdG}
\mathcal{H}_{\mathrm{BdG}}^{\bm{k},\bm{q}}=\mathcal{H}_{\mathrm{BdG}}^{\bm{k}}+q_a \hat{v}_a /2+q_a q_b\hat{w}_{ab}/8+\mathcal{O}(\bm{q}^3),
\end{equation}
where the velocity operators are 
\begin{equation}
\hat{v}_a=\left(\begin{array}{cc}
 \hat{\mathcal{V}}_{a}(\bm{k})& 0 \\
0 & -\hat{\mathcal{V}}^*_{a}(-\bm{k})
\end{array}\right),
\hat{w}_{ab}=\left(\begin{array}{cc}
 \mathcal{W}_{ab}(\bm{k}) & 0 \\
0 & -\mathcal{W}_{ab}^*(-\bm{k})
\end{array}\right).
\end{equation}
Here we introduce $\mathcal{V}_a(\bm{k})=\partial H_{\bm{k}}/\partial k_a$ and $\mathcal{W}_{ab}(\bm{k})=\partial^2 H_{\bm{k}}/\partial k_a \partial k_b$. 
We proceed in evaluating terms of Eq.~(\ref{eq:HBdG}), we note that if $A$ and $B$ are matrices and $B\ll A$ ($B\sim \mathcal{O}(\bm{q})$), we can expand the matrix expression in small $B$ as 
\begin{equation}
\label{eq:expandmatrix}
(A+B)^{-1}=A^{-1}-A^{-1}B A^{-1}+A^{-1}B A^{-1}B A^{-1}+\mathcal{O}(B^3).
\end{equation}
In the following we will apply Eq.~(\ref{eq:expandmatrix}) to 
the full Green function in order to make an effective expansion around small $\bm{q}$. We obtain 
\begin{equation}
\label{eq:green_expand}
\mathcal{G}(\bm{k},\bm{q},i\omega_n)=(i\omega_n-\mathcal{H}_{\mathrm{BdG}}^{\bm{k},\bm{q}})^{-1}=\mathcal{G}_0+\frac{1}{2}q_a \mathcal{G}_0 \hat{v}_a \mathcal{G}_0+q_a q_b (\frac{1}{8}\mathcal{G}_0 \hat{w}_{ab}\mathcal{G}_0+\frac{1}{4}\mathcal{G}_0 \hat{v}_a \mathcal{G}_0 \hat{v}_b \mathcal{G}_0)
\end{equation}
Here $\mathcal{G}_0 \equiv \mathcal{G}(\bm{k},i\omega_n)=(i\omega_n-\mathcal{H}_{\mathrm{BdG}}^{\bm{k}})^{-1}$. We insert Eq.~(\ref{eq:green_expand}) to Eq.~(\ref{eq:magnetization}) to get the magnetization susceptibility at both linear- and (nonlinear) second-order. Firstly, the first-order spin susceptibility $\alpha_{ab}$ is described by $\delta M_a^{(1)}=\alpha_{ab} q_b$, where $\alpha_{ab}$ reads
\begin{equation}
\label{eq:linear_sus_supple}
\alpha_{ab}=-\frac{g_s\mu_B k_B T}{8}\sum_{n\bm{k}}\mathrm{Tr}[\eta_a \mathcal{G}_0 \hat{v}_b \mathcal{G}_0].
\end{equation}
More importantly, we can write the second-order nonlinear spin magnetization along $c$ direction as $\delta M_c^{(2)}=\chi_{ab}^c q_a q_b$, and the nonlinaer susceptibility $\chi$ can be obtained as
\begin{equation}
\label{eq:nonlinear_sus_supple}
\chi_{ab}^c=-\frac{g_s\mu_B k_B T}{32}\!\sum_{n\bm{k}}\!\mathrm{Tr}[\eta_c(\mathcal{G}_0\hat{w}_{ab} \mathcal{G}_0+2\mathcal{G}_0\hat{v}_{a}\mathcal{G}_0\hat{v}_{b}\mathcal{G}_0)].
\end{equation}
\subsection{Analytical result}
In the previous subsection, we use the Green's function method to give the general formulas of the linear and nonlinear magnetization. These formulas can be adopted to evaluate the spin susceptibilities for altermagnet/superconductor heterostructures. In this subsection, we provide the analytical solution of the nonlinear spin susceptibility $\chi^z_{ab}$. For ideal altermagnets, we can write down a general two-band model
\begin{equation}
\label{eq:altermagnet_full}
H_{\bm{k}}=[ta^2(k_x^2+k_y^2)-\mu]s_0+J_{ex}a^2[2k_xk_y\cos(2\theta)+(k_x^2-k_y^2)\sin(2\theta)]s_z,
\end{equation}
where $t$ parameterizes the usual kinetic energy, $J_{ex}$ denotes the $d$-wave exchange magnetic order parameter and $\mu$ is the chemical potential. $\theta$ is the angle between the altermagnetic orientation and the crystalline axes. The velocity operators are 
\begin{equation}
\hat{v}_x=a^2\left(\begin{array}{cc}
 2tk_xs_0+2J_{ex}k\sin(\phi+2\theta)s_z & 0 \\
0 & 2tk_xs_0+2J_{ex}k\sin(\phi+2\theta)s_z
\end{array}\right),
\end{equation}
\begin{equation}
\hat{v}_y=a^2\left(\begin{array}{cc}
 2tk_ys_0+2J_{ex}k\cos(\phi+2\theta)s_z & 0 \\
0 & 2tk_ys_0+2J_{ex}k\cos(\phi+2\theta)s_z
\end{array}\right),
\end{equation}
and
\begin{equation}
\hat{w}_{xx}=a^2\left(\begin{array}{cc}
 2ts_0+2J_{ex}\sin(2\theta)s_z & 0 \\
0 & -2ts_0-2J_{ex}\sin(2\theta)s_z
\end{array}\right),
\hat{w}_{xy}=a^2\left(\begin{array}{cc}
 2J_{ex}\cos(2\theta)s_z & 0 \\
0 & -2J_{ex}\cos(2\theta)s_z
\end{array}\right).
\end{equation}
Here $k_x=k\cos(\phi)$ and $k_y=k\sin(\phi)$.
For $J_{ex}\ll t$, we can obtain an approximate analytical results 
as $\chi^z_{xx}=\mu_B\sin(2\theta)\int_0^\infty f(x)dx+ \mathcal{O}(J_{\rm ex}^2)$ and $\chi^z_{xy}=\chi^z_{yx}=\mu_B\cos(2\theta)\int_0^\infty f(x)dx+ \mathcal{O}(J_{\rm ex}^2)$, where
\begin{equation}
\label{eq:analytical}
f(x)=\frac{\beta^3 J_{\mathrm{ex}}}{64\pi} x\mathrm{sech}^2\gamma \Big[\frac{4}{\beta^2}+t^2x^4(2-3\mathrm{sech}^2\gamma)+\frac{4tx^2(\mu-tx^2)\tanh\gamma}{\gamma}\Big],
\end{equation}
where $\gamma=\beta\sqrt{\Delta^2+(tx^2-\mu)^2}/2$.  
\begin{figure}
		\centering
		\includegraphics[width=1\linewidth]{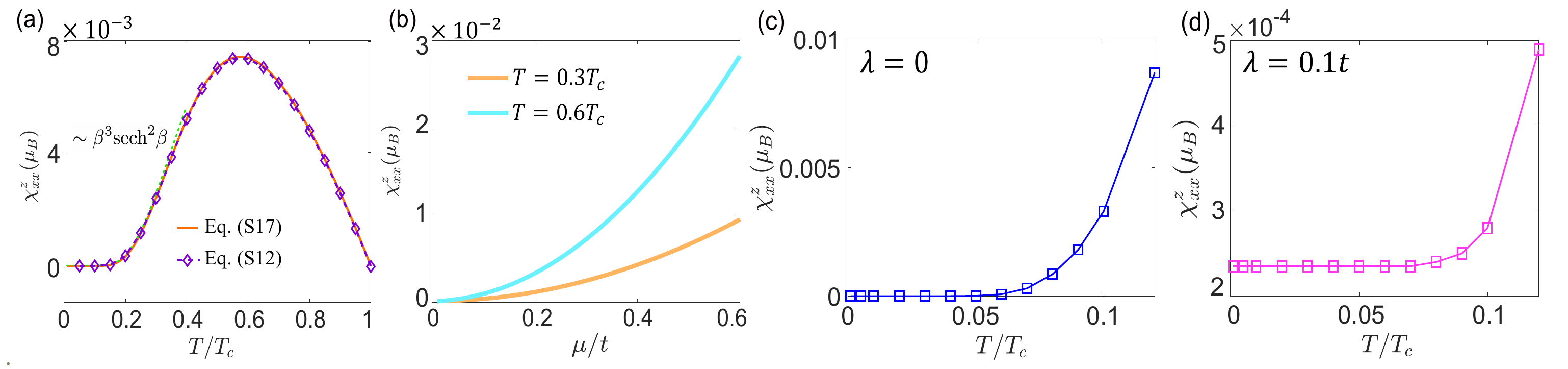}
		\caption{(a) The second-order spin susceptibility $\chi_{xx}^z$ as a function of the temperature $T$. The purple diamonds denotes the numerical results from Eq.~\eqref{eq:nonlinear_sus_supple} and the orange line denotes the analytical result from Eq.~\eqref{eq:analytical}. At low temperatures, $\chi^z_{xx}\sim \beta^3\mathrm{sech}^2\beta$. (b) $\chi_{xx}^z$ as a function of the Fermi energy $\mu$ at $T=0.3,0.6T_c$. (c),(d) The calculations of $\chi_{xx}^z$ for $\lambda=0$ (c) and $\lambda=0.1t$ (d). In (c), $\chi_{xx}^z$ is suppressed as $T\rightarrow 0$ because of the factor $\frac{1}{T^3}\mathrm{sech}^2(\frac{1}{T})$. In (d), $\chi_{xx}^z$ saturate to a finite value as $T\rightarrow 0$ for nonzero SOI. Other Parameters: $J=0.02t$ for (a) and (b). $J=0.2t$ for (c) and (d). $\mu=0.3t, \Delta_0=0.1t$ for all four panels.}
		\label{fig:figS1}
\end{figure}

By employing Eq.~\eqref{eq:analytical}, we plot $\chi_{xx}^z$ as a function of temperature in Fig.~\ref{fig:figS1}(a) (orange solid curve) with $J_{\mathrm{ex}}=0.02t$. By way of comparison, we plot a numerical evaluation of Eq.~\eqref{eq:nonlinear_sus_supple} (purple diamonds); both agree with each other. At low temperatures ($\beta \Delta_0\gg 1$), $\chi_{xx}^z$ starts to grows with $T$ scaling as $\chi_{xx}\sim \beta^3\mathrm{sech}^2\beta$. Moreover, at $T/T_c\sim 0.6$, $\chi_{xx}^z$ reaches a peak after which it rapidly diminishes with temperature. This non-monotonic behavior contrasts sharply with that of the linear magnetoelectric effect in Rashba-type superconductors~\cite{he2019spin} that instead saturates at low temperature (see discussion below). We also plot the $\mu$ dependence of $\chi_{xx}^z$ from Eq.~\eqref{eq:analytical} in Fig.~\ref{fig:figS1}(b). Larger NSM response can be achieved by increasing the Fermi energy $\mu$ because of the enhanced density of states.

From this analytical formula, we find that at low temperature, $\chi$ is suppressed as $\frac{1}{T^3}\mathrm{sech}^2(\frac{1}{T})$. This can be understood in the following way: for an ideal altermagnet ($\lambda=0$), $\eta_z$ is conserved (recall $\eta_z=\mathrm{diag}[s_z,-s_z]$), namely $[\eta_z,\mathcal{H}_{\mathrm{BdG}}^{\bm{k},\bm{q}}]=0$, which indicates that nonzero $M_z$ arises from the thermal fluctuations in the current-carrying state at nonzero temperature. The suppression of $\chi_{xx}^z$ is shown in Fig.~\ref{fig:figS1}(c) with $\lambda=0$ (zero SOI). In the presence of SOI ($\lambda\neq 0$), $[\eta_z,\mathcal{H}_{\mathrm{BdG}}^{\bm{k},\bm{q}}]\neq 0$, leading to finite $M_z$ even when $T\rightarrow 0$. 
This is shown in Fig.~\ref{fig:figS1}(d). 

\subsection{Effects from local mirror breaking}
When fabricating a hybrid magnetic material/superconductor heterostructure, surface reconstruction at the interface can ensue. For instance, surface reconstruction at the interface can break the mirror symmetry locally rendering the total system effectively noncentrosymmetric~\cite{linder2011spin}. This yields an interfacial spin orbital interaction (SOI) $H_{\mathrm{SOI}}=V_{s}(\bm{k}\times \bm{\sigma})\cdot \hat{\bm{n}}$ where $\hat{\bm{n}}$ is an interface normal vector. When $\hat{\bm{n}}=\hat{z}$, $V_{s}$ describes the well-known Rashba-like SOI, which preserves $C_4\mathcal{T}$. When $\hat{\bm{n}}=\hat{x}$ or $\hat{y}$, $k_y\sigma_z$ or $k_x\sigma_z$ can emerge as additional terms which break $C_4\mathcal{T}$. Since we have studied the Rashba SOI in the main text and above, here we present a simple model to illustrate the effect from local mirror symmetry breaking. To be specific, we consider the following effective Bloch Hamiltonian
 \begin{equation}
\label{eq:rashba_hamil}
H_{\bm{k}}=[ta^2(k_x^2+k_y^2)-\mu]s_0+\lambda a(k_x s_y-k_y s_x)+J_{\mathrm{ex}}a^2(k_x^2-k_y^2)s_z+(\gamma_x k_x+\gamma_y k_y)a\sigma_z,
\end{equation}
where $\gamma_x$ and $\gamma_y$ denote the mirror breaking terms. 

\begin{figure}
		\centering
		\includegraphics[width=0.7\linewidth]{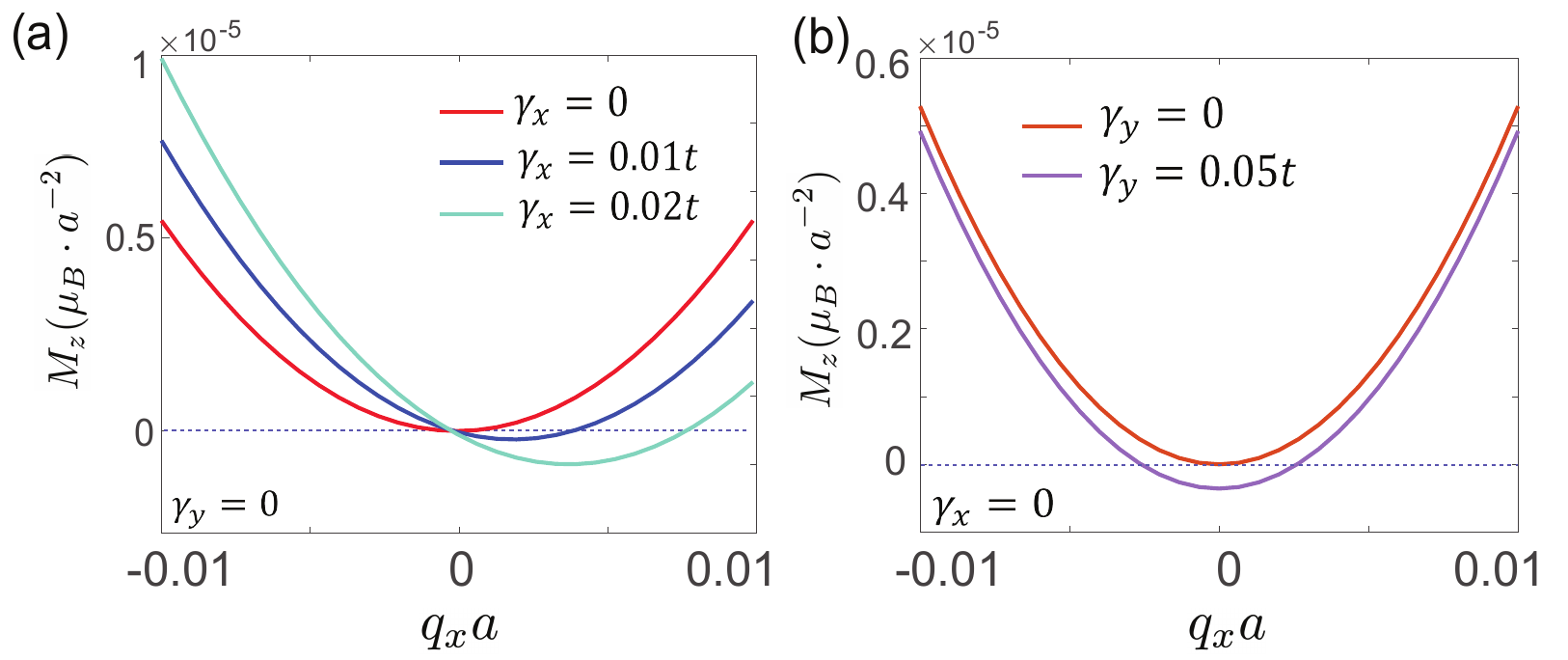}
		\caption{NSM with local mirror symmetry breaking from interfacial spin-orbital interaction. In (a) $\gamma_y=0$ and in (b) $\gamma_x=0$.
 $\gamma_x,\gamma_y$ denotes the strength of SOI for the mirror breaking. Parameters: $J=0.2t, \mu=0.3t, \lambda=0.05t, \Delta_0=0.1t$ and $T=0.3T_c$.}
		\label{fig:figS2}
\end{figure}

In Fig.~\ref{fig:figS2}(a), we plat $M_z$ with finite $\gamma_x$ and $\gamma_y=0$. It is clear that in the presence of $\gamma_x$, we have a general form for the current induced magnetization as $M_z=Aq_x^2+Bq_x+C$. The linear term parameterized by the coefficient B arises from nonzero $\gamma_x$ which breaks mirror symmetry $\mathcal{M}_y$. In Fig.~\ref{fig:figS2}(b), we plot $M_z$ with finite $\gamma_y$ and $\gamma_x=0$. The $\gamma_y$ term breaks mirror symmetry $\mathcal{M}_x$ but preserve $\mathcal{M}_y$, which forbids the linear term B. Note that in both cases, we have $M_z\neq0$ at $q_x=0$, resulting from the fact that $C_4 \mathcal{T}$ is nor preserved in the presence of $\gamma_x$ or $\gamma_y$. Importantly, in all cases, NSM is robust and can dominate the magnetoelectric response. Interestingly, the combination of both linear and nonlinear superconducting magnetoelectric effects in Fig.~\ref{fig:figS2}(a) leads to a {\it non-reciprocal} magneto-electric effect 
with magnetizations distinct for $+q$ and $-q$. 

\begin{figure}
		\centering
		\includegraphics[width=1\linewidth]{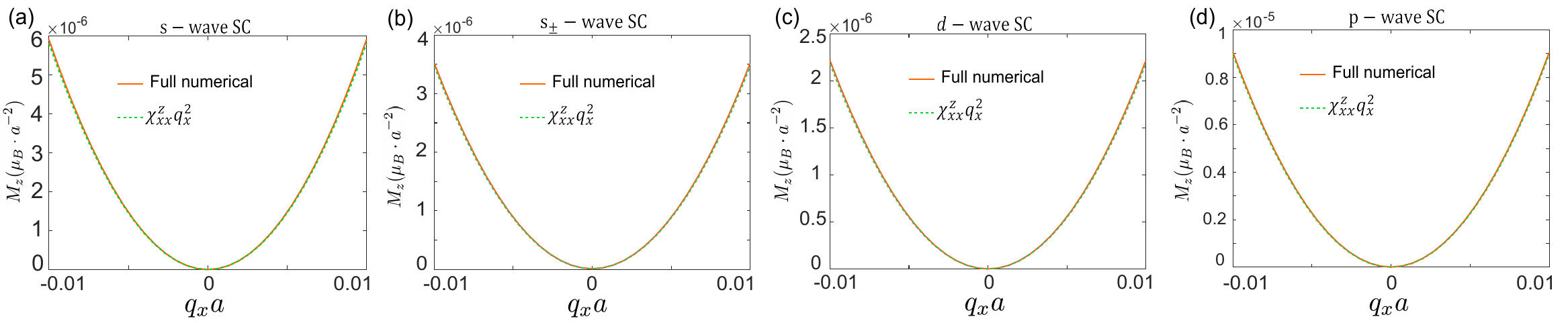}
		\caption{The supercurrent induced magnetization for (a) Spin-singlet $s$-wave pairing with $\hat{\Delta}_{\bm{k}}=\Delta_0is_y$ at $\Delta_0=0.1t$, (b) Spin-singlet $s_{\pm}$-wave pairing with $\hat{\Delta}_{\bm{k}}=(\Delta_0+\Delta_1\bm{k}^2)is_y$ at $\Delta_0=0.1t$ and $\Delta_1=0.05t$, (c) Spin-singlet $d$-wave pairing with $\hat{\Delta}_{\bm{k}}=\Delta_0(k_x^2-k_y^2)is_y$ at $\Delta_0=0.1t$, (d) Spin-triplet chiral $p$-wave pairing with $\hat{\Delta}_{\bm{k}}=\Delta_0[(k_x+ik_y)s_z]is_y$ at $\Delta_0=0.1t$. Other parameters: $(J_{\mathrm{ex}},\lambda,\mu)=(0.2,0.05,0.3)t$. The temperature is set to be $k_B T=0.2\Delta_0$ for all four panels.}
		\label{fig:figS3}
\end{figure}

\subsection{NSM with unconventional pairing states}
In the main text, we focus on the nonlinear magnetic response with spin-singlet $s$-wave pairing. Here we present more examples for the NSM with different unconventional pairing states. We consider (1) Spin-singlet $s_{\pm}$-wave pairing with $\hat{\Delta}_{\bm{k}}=(\Delta_0+\Delta_1\bm{k}^2)is_y$. (2) Spin-singlet $d$-wave pairing with $\hat{\Delta}_{\bm{k}}=\Delta_0(k_x^2-k_y^2)is_y$. (3) Spin-triplet $p$-wave pairing with $\hat{\Delta}_{\bm{k}}=\Delta_0[(k_x+ik_y)s_z]is_y$. For simplicity, we consider the cases that the pairing states do not break the $C_4\mathcal{T}$ symmetry of $H_{\mathrm{BdG}}$. 

In Fig.~\ref{fig:figS3} we calculate the $M_z$ for these different pairing potentials. For these different superconducting pairing states, $M_z$ can be quantitatively different, but $M_z \sim q_x^2$ persists as a general behavior (quadratic in Cooper pair momentum).

\subsection{Details of material calculation}
\begin{figure}
		\centering
		\includegraphics[width=0.8\linewidth]{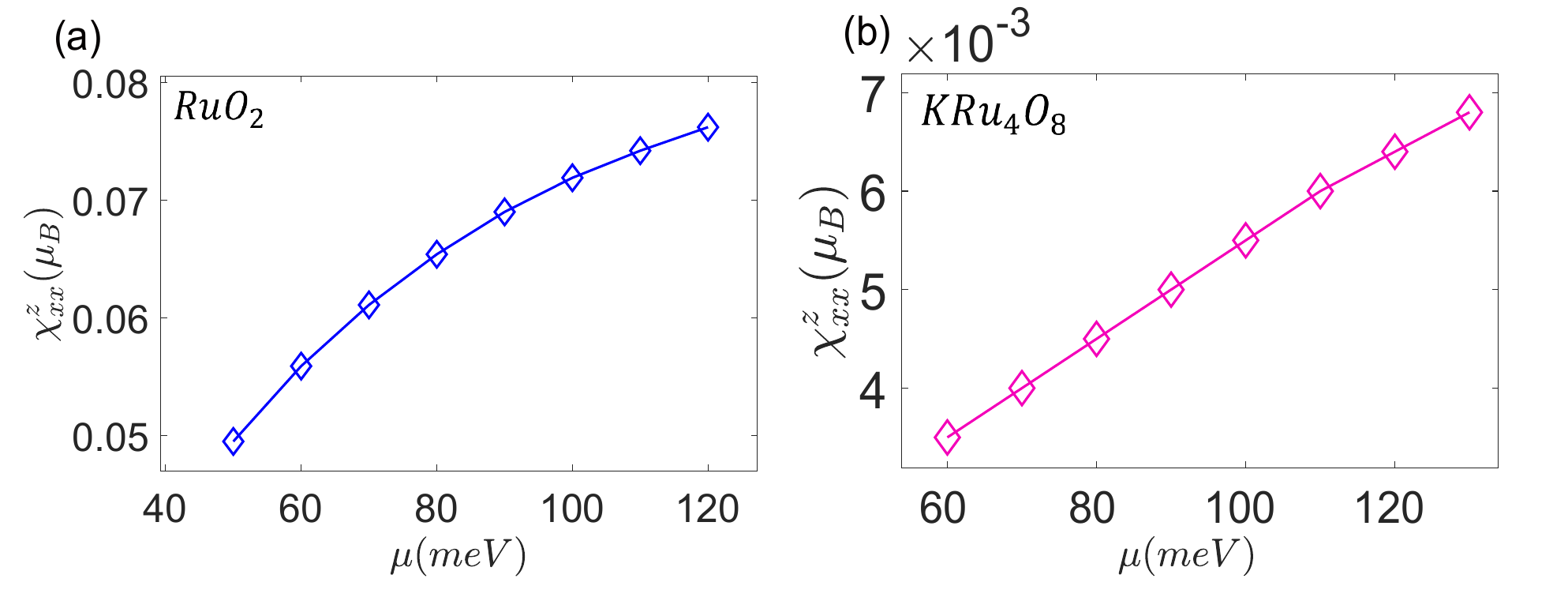}
		\caption{The calculation of $\chi_{xx}^z$ [Eq.~\eqref{eq:nonlinear_sus_supple}] for RuO$_2$ (a) and KRu$_4$O$_8$ (b) as a function of Fermi energy $\mu$. For RuO$_2$, $t=2.5$ eV, $J_{\mathrm{ex}}=0.5$ eV, $\lambda=40$ meV. For KRu$_4$O$_8$, $t=0.05$ eV, $J_{\mathrm{ex}}=0.018$ eV, $\lambda=40$ meV. We take $\Delta_0=1$meV and $T=0.3T_c$ for the two plots.}
		\label{fig:figS4}
\end{figure}

Here we provide more details of the two examples in the main text. For RuO$_2$ with in-plane lattice constant $a=4.5\AA$, and $t=2.5$ eV, $J_{\mathrm{ex}}=0.5$ eV, we assume the Rashba SOI term is $\lambda=40$ meV, which is an estimation including both the intrinsic and interface contributions. Using a pairing gap of $\Delta_0=1$ meV and $T=0.3T_c$ we plot $\chi_{xx}^z$ as a function of Fermi energy $\mu$ in Fig.\ref{fig:figS4}(a). Larger response for larger $\mu$ is found because of the enhanced density of states. For KRu$_4$O$_8$ we adopt the parameters: $t=0.05$ eV, $J_{\mathrm{ex}}=0.018$ eV, and $a=9.9\AA$. The SOI and pairing gap is the same as RuO$_2$, and we also get a similar result in Fig.~\ref{fig:figS4}(b). 

We take RuO$_2$ as an example to estimate the magnetization. For a Cooper pair momentum $q\approx  10^{-3}\AA^{-1}$, the induced out-of-plane spin density in the RuO$_2$ thin film/superconductor heterostructure is $\sim 10^{-5}\mu_B/\mathrm{nm}^2$ at $\chi_{xx}^z=0.1\mu_B$. Considering a thin film with the lattice constant $c=0.31$nm, the magnetization in the standard unit is $\delta M=3\times 10^{-5}\mu_B/\mathrm{nm}^3$. Using this, we estimate the equivalent magnetic field as $\delta B=\mu_0\delta M\approx 0.4\times 10^{-6}$T.

\section{Proximity effect and pairing correlations} 

Although  we only consider the simplest conventional spin singlet pairing, i.e., $\Delta_0 is_y$, here we show that the $d$-wave magnetic order will induce the spin-triplet correlations. To show this, the BdG Hamiltonian is
\begin{equation}
	\mathcal{H}_{\mathrm{BdG}}^{\bm{k}}=\begin{pmatrix}
		H_{\bm{k}}& \Delta_0 is_y\\
		(\Delta_0 is_y)^{\dagger}& -H^*_{-\bm{k}}
	\end{pmatrix}\label{noorbit}
\end{equation}
Here $H_{\bm{k}}$ is from Eq.\eqref{eq:altermagnet_full}. Let us identify the superconducting properties in terms of Green's function:
\begin{eqnarray}
G_{\lambda\mu}(\bm{k},\tau)&=T_{\tau}\{c_{\bm{k},\lambda}(\tau)c^{\dagger}_{\bm{k},\mu}(0)\},\\
F_{\lambda\mu}(\bm{k},\tau)&=T_{\tau}\{c_{\bm{k},\lambda}(\tau)c_{-\bm{k},\mu}(0)\}.
\end{eqnarray}
We can rewrite the Green's function in the Matsubara frequency space: $G_{\lambda\mu}(\bm{k},i\omega_n)=\int_0^{\beta}d\tau e^{i\omega_n\tau}G_{\lambda\mu}(\bm{k},\tau)$ and $F_{\lambda\mu}(\bm{k},i\omega_n)=\int_0^{\beta}d\tau e^{i\omega_n\tau}F_{\lambda\mu}(\bm{k},\tau)$. The latter $F_{\lambda\mu}(\bm{k},i\omega_n)$ represents the pairing correlations we refer.  These two Green's functions are related to the Gor'kov Green's function as
\begin{equation}
\label{eq:green_full}
	\mathcal{G}(\bm{k},i
	\omega_n)=(i\omega_n-H_{\mathrm{BdG}}^{\bm{k}})^{-1}=\begin{pmatrix}
		G_e(\bm{k},i\omega_n)&F(\bm{k},i\omega_n)\\F^{\dagger}(\bm{k},i\omega_n)&G_h(\bm{k},i\omega_n)
	\end{pmatrix}.
\end{equation}
Substituting the BdG Hamiltonian into Eq.~\eqref{eq:green_full} and after some algebraic manipulation, we can parameterize the pairing correlation as 
\begin{equation}
F(\bm{k},i\omega_n)=\Delta_0[C_1(\bm{k},i\omega_n)+C_2(\bm{k},i\omega_n)\bm{d}(\bm{k})\cdot \bm{s}]is_y \label{pairing _corr}
\end{equation}
with the coefficients 
\begin{equation}
	C_1(\bm{k},i\omega_n)=-\frac{1}{2}[\frac{1}{\Delta_0^2+\xi_k^2-(J_{ex} f_k^2+i\omega_n)^2}+\frac{1}{\Delta_0^2+\xi_k^2-(J_{ex} f_k^2-i\omega_n)^2}],
\end{equation}
\begin{equation}
	C_2(\bm{k},i\omega_n)=\frac{2i\omega_n}{(\Delta_0^2+\xi_k^2-J_{ex}^2f_k^2)^2+2(\Delta_0^2+\xi_k^2+J_{ex}^2f_k^2)\omega_n^2+\omega_n^4},
\end{equation}
where $\xi_k=ta^2(k_x^2+k_y^2)-\mu$ and $f_k=a^2[2k_xk_y\cos(2\theta)+(k_x^2-k_y^2)\sin(2\theta)]$.
Importantly, the triplet vector is directly related to the $d$-wave magnetic order with 
\begin{equation}
	\bm{d}(\bm{k})= (0,0,J_{ex}f_k),
\end{equation}
Therefore, it can be seen that due to the presence of $d$-wave, the spin-singlet pairing and the spin-triplet pairing are mixed. Clearly, $C_2(\bm{k},i\omega_n)=-C_2(-\bm{k},-i\omega_n)$, which implies that the pairing amplitude changes sign under time-reversal operation. Therefore, the spin-triplet pairing is a type of odd-frequency pairing state~\cite{linder2019odd}.

\section{Linear Edelstein effect in Rashba-type superconductors}
In this section, we review the linear magnetoelectric effect in superconductors with Rashba SOC. The low energy Bloch Hamiltonian is 
\begin{equation}
\label{eq:rashba_hamil}
H_{\bm{k}}=[ta^2(k_x^2+k_y^2)-\mu]s_0+\lambda a(k_x s_y-k_y s_x),
\end{equation}
where $\lambda$ characterizes the strength of Rashba SOI, and $a$ is the lattice constant. We introduce the spin singlet pairing $\Delta_{\bm{k}}=\Delta_0 is_y$. Similarly, the linear spin susceptibility $\alpha_{yx}$ can be obtained as
\begin{equation}
\alpha_{yx}=\mu_B\int_{0}^\infty f(x)dx
\end{equation}
where
\begin{equation}
\label{eq:rashbame_ana}
\begin{split}
f(x)=&\frac{\beta}{32\pi(t x^2-\mu)}\{ x(\lambda-2tx)(tx^2-\mu)\mathrm{sech}^2\gamma_1+x(\lambda+2tx)(tx^2-\mu)\mathrm{sech}^2\gamma_2\\
&-[\Delta^2+(\mu-t x^2)(\mu+\lambda x -tx^2)]\tanh\gamma_1/\gamma_1+[\Delta^2+(\mu-t x^2)(\mu-\lambda x -tx^2)]\tanh\gamma_2/\gamma_2\}
\end{split}
\end{equation}
with
\begin{equation}
\gamma_1 =\frac{\beta\sqrt{\Delta^2+(\mu+\lambda x-t x^2)^2}}{2},
\gamma_2 =\frac{\beta\sqrt{\Delta^2+(\mu-\lambda x-t x^2)^2}}{2}
\end{equation}
\begin{figure}
		\centering
		\includegraphics[width=1\linewidth]{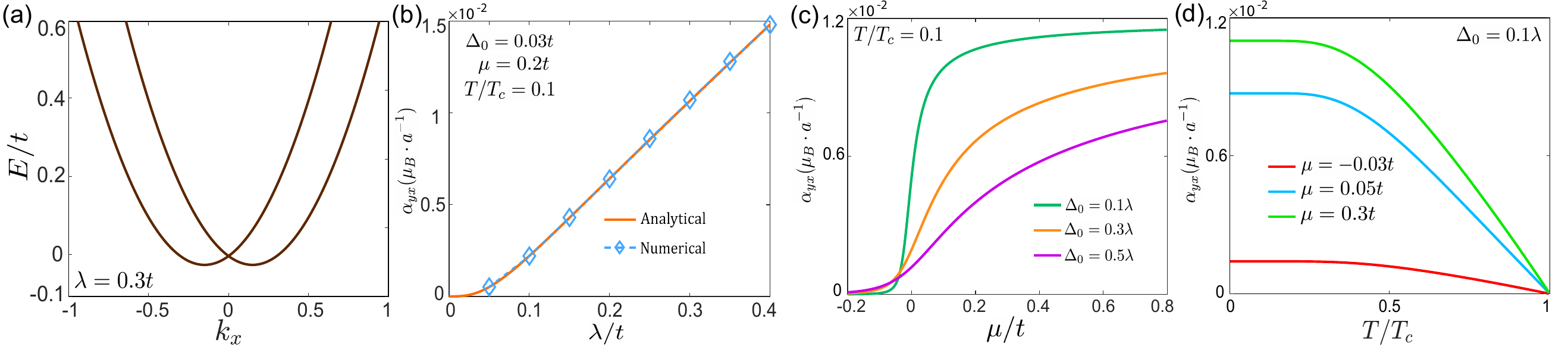}
		\caption{The linear spin magnetization in Rashba-type superconductors. (a) The band structure with Rashba SOC at $\lambda=0.3t$ from Eq.~\eqref{eq:rashba_hamil}. (b) The first-order spin susceptibility $\alpha_{yx}$ as a function of $\lambda$ at $T=0.1T_c$, $\Delta_0=0.03t$ and $\mu=0.2t$. The purple diamonds denotes the numerical results from Eq.~\eqref{eq:linear_sus_supple} and the orange line denotes the analytical result from Eq.~\eqref{eq:rashbame_ana}. (c) The first-order spin susceptibility $\alpha_{yx}$ as a function of $\mu$ at $T=0.1T_c$ for various values of the pairing gap $\Delta_0$. (d) $\alpha_{yx}$ as a function of $T$ at $\Delta_0=0.1\lambda$ for various values of $\mu$.}
		\label{fig:figS5}
\end{figure}

We apply the above formula to compute the linear (first order) magnetization in superconductors with Rashba SOI. The calculated band structure from Eq.~\eqref{eq:rashba_hamil} is shown in Fig.~\ref{fig:figS5}(a) at $\lambda=0.3t$. By employing Eq.~\eqref{eq:rashbame_ana}, we plot $\alpha_{yx}$ as a function of $\lambda$ in Fig.~\ref{fig:figS5}(b) (orange solid curve). By way of comparison, we also show a numerical plot of Eq.~\eqref{eq:linear_sus_supple} (blue diamonds); both agree with each other. In Fig.~\ref{fig:figS5}(c) we show the $\mu$-dependence of $\alpha_{yx}$ with various values of $\Delta_0$. When $\mu<0$, $\alpha_{yx}$ decreases as $\mu$ goes down. Fig.~\ref{fig:figS5}(d) shows $\alpha_{yx}$ as a function of temperature for different values of $\mu$. Generally, $\alpha_{yx}$ starts to grow linearly when $T<T_c$, and it saturates when $T\rightarrow 0$. This behavior is consistent with the results shown in Ref.~\cite{he2019spin}.

\section{Symmetry analysis of NSM effect}

\begin{table}
\caption{Symmetry restrictions of the NSM tensor $\chi_{ab}^c$ in a two-dimensional system. We have symmetrized $\chi_{xy}^c$ by defining $\chi_{(xy)}^c\equiv (\chi_{xy}^c+\chi_{yx}^c)/2$. Here $\checkmark$ indicates allowed and $\times$ indicates forbidden. Here the $/$ symbol denotes ``or''.}
\label{tab:tensor}
\begin{ruledtabular}
\begin{tabular}{l|llllllllllll}
 & $P$ & $C_{2x}/\mathcal{M}_x$ & $C_{2x}\mathcal{T}/\mathcal{M}_x \mathcal{T}$ & $C_{2y}/\mathcal{M}_y$ & $C_{2y}\mathcal{T}/\mathcal{M}_y \mathcal{T}$ & $C_{2z}/\mathcal{M}_z$ & $C_{2z}\mathcal{T}/\mathcal{M}_z \mathcal{T}$ & $C_{3z}$ & $C_{4z}/C_{6z}$ & $C_{4z}\mathcal{T}$ & $C_{6z}\mathcal{T}$ & $C_{8z}\mathcal{T}/C_{12z}\mathcal{T}$\\
 \hline
$\chi_{xx}^x$ & \checkmark & \checkmark    & $\times$  & $\times$     & \checkmark &   $\times$ & \checkmark & \checkmark     & $\times$ &   $\times$ & \checkmark & $\times$   \\
$\chi_{(xy)}^x$ & \checkmark & $\times$     & \checkmark & \checkmark     & $\times$  &   $\times$ & \checkmark & $-\chi_{yy}^y$ & $\times$ &   $\times$ & $-\chi_{yy}^y$ & $\times$  \\
$\chi_{yy}^x$ & \checkmark & \checkmark     & $\times$  &  $\times$     & \checkmark &   $\times$ & \checkmark & $-\chi_{xx}^x$ & $\times$ &   $\times$ & $-\chi_{xx}^x$ & $\times$  \\
$\chi_{xx}^y$ & \checkmark & $\times$     & \checkmark & \checkmark     & $\times$ &   $\times$ & \checkmark & $-\chi_{yy}^y$ & $\times$ &   $\times$ & $-\chi_{yy}^y$  & $\times$ \\
$\chi_{(xy)}^y$ & \checkmark & \checkmark     & $\times$  & $\times$   & \checkmark &   $\times$ & \checkmark & $-\chi_{xx}^x$ & $\times$ &   $\times$ & $-\chi_{xx}^x$ &  $\times$  \\
$\chi_{yy}^y$ & \checkmark & $\times$     & \checkmark &  \checkmark     & $\times$ &   $\times$ & \checkmark & \checkmark     & $\times$ &   $\times$ & \checkmark & $\times$ \\
$\chi_{xx}^z$ & \checkmark & $\times$      & \checkmark  & $\times$     & \checkmark &   \checkmark & $\times$ & \checkmark     & \checkmark &   \checkmark & $\times$ & $\times$  \\
$\chi_{(xy)}^z$ & \checkmark & \checkmark    & $\times$ &  \checkmark    & $\times$ &   \checkmark & $\times$ & $\times$ & $\times$ &   \checkmark & $\times$ & $\times$  \\
$\chi_{yy}^z$ & \checkmark & $\times$     & \checkmark  &  $\times$     & \checkmark &   \checkmark & $\times$ & $\chi_{xx}^z$ & $\chi_{xx}^z$ &   $-\chi_{xx}^z$ & $\times$ & $\times$  
\end{tabular}
\end{ruledtabular}
\end{table}

\subsection{Symmetry analysis}
In this section, we study the symmetry requirements for NSM (i.e. obtaining a non-zero $\chi_{ab}^c$) focussing on systems with time-reversal $\mathcal{T}$ breaking. 
Under a (composite) operation 
$\mathcal{T}^{s_{\mathcal{T}}}\mathcal{R}$, the NSM tensor $\chi_{ab}^c$ transforms as a
third-rank pseudotensor, which can be written as
\begin{equation}
\chi_{a'b'}^{c'}=(-1)^{s_{\mathcal{T}}}\mathrm{det}(\mathcal{R})\mathcal{R}_{c'c}\mathcal{R}_{b'b}\mathcal{R}_{a'a}\chi_{ab}^{c}
\end{equation}
where $\mathcal{R}$ is the point group operation. Here we have used a short form notation to indicate composite operation. For point group $\mathcal{R}$ operation only, $s_{\mathcal{T}}=0$. For composite $\mathcal{R}\mathcal{T}$ operation, $s_{\mathcal{T}}=1$. We examine the symmetry requirements of NSM, listing how various (composite) point group symmetries affect the presence/absence of NSM as shown 
in Table.~\ref{tab:tensor}. 
From this symmetry analysis, we can clearly see that NSM can arise for a range of symmetries beyond $C_{4}\mathcal{T}$ that we focus on in the main text. 

It is also worthwhile to examine the presence/absence of NSM response for other types of altermagnetic order. For example, $g$-wave altermagnetic order $J_{ex}k_x k_y(k_x^2-k_y^2)s_z$ and $i$-wave altermagnetic order $J_{ex}k_x k_y(3k_x^2-k_y^2)(3k_y^2-k_x^2)s_z$~\cite{vsmejkal2022beyond}. It is clear that for altermagnetic $g$-wave with $C_{8z}\mathcal{T}$ symmetry or $i$-wave with $C_{12z}\mathcal{T}$ symmetry, $\chi_{ab}^c=0$. 

\subsection{A case study: NSM in kagome noncollinear antiferromagnet}
In the main text, we focussed on the nonlinear superconducting magnetoelectric (NSM) effect in $d$-wave altermagnet/superconductor heterostructure. The formula we derive in Eq.~\eqref{eq:nonlinear_sus_supple} is general and can also be applied to other magnet/superconductor heterostructure or intrinsic superconductor with magnetism. Indeed, in the symmetry analysis of the previous section, we find that NSM can appear for a range of other symmetries. In this section we present more details of another example (briefly described in the main text): Noncollinear antiferromagnet with $D_{3d}$ symmetry on kagome lattice~\cite{lee2024fermi}. 

\begin{figure}
		\centering
		\includegraphics[width=1\linewidth]{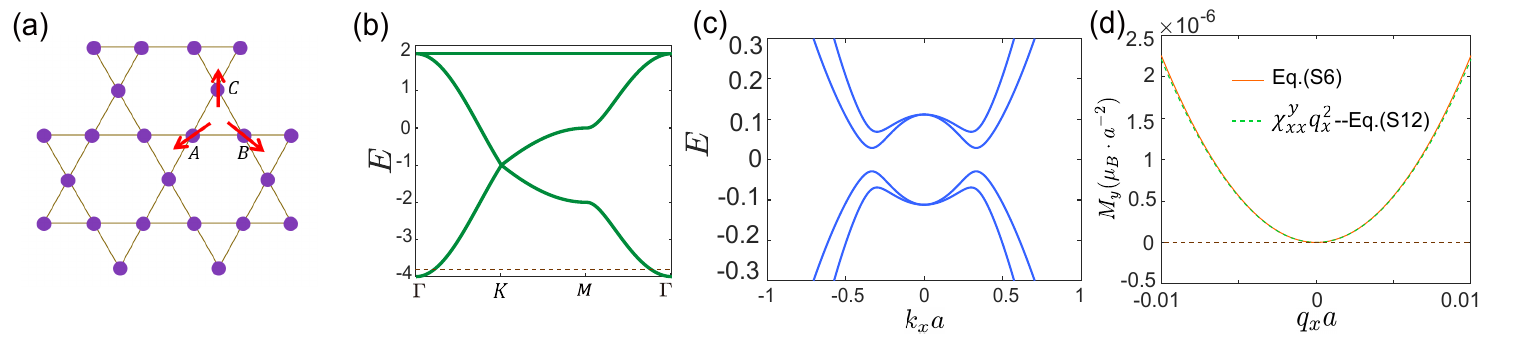}
		\caption{(a) The crystal structure for the kagome lattice. The noncollinear magnetic order is labeled by red arrows. (b) The energy spectrum of kagome lattice from Eq.~\eqref{eq:kagome_hamil}. (c) The BdG spectrum of the kagome oncollinear antiferromagnet [Eq.~\eqref{eq:model_kagome}] with the pairing potential $\Delta_0 is_y$. (d) The spin magnetization $M_y$ as a function of $q_x$ from full numerical method in Eq.~\eqref{eq:magnetization} (solid orange). As a comparison, the NSM susceptibility multiplied by $q_x^2$ is plotted as Eq.~\eqref{eq:nonlinear_sus_supple} (dashed green). Parameters in the calculation:$t=1,m=0.2,\mu=0.1,\Delta_0=0.05$ and $T=0.3T_c$.}
		\label{fig:figS6}
\end{figure}
Considering a kagome lattice with $A,B,C$ atoms per uint cell, in the basis $c_{\bm{k}}^\dagger=(c_{A\bm{k}}^\dagger,c_{B\bm{k}}^\dagger,c_{C\bm{k}}^\dagger)$ the tight binding Hamiltonian is given by
\begin{equation}
\label{eq:kagome_hamil}
	H_{kagome}(\bm{k})=\begin{pmatrix}
		0& h_{12} & h_{13}\\
		h_{12}^* & 0 & h_{23}\\
        h_{13}^* & h_{23}^* & 0
	\end{pmatrix}
\end{equation}
where $h_{12}=-(1+e^{i\bm{k}\cdot \bm{a_1}})$, $h_{13}=-(1+e^{-i\bm{k}\cdot \bm{a_3}})$, $h_{23}=-(1+e^{i\bm{k}\cdot \bm{a_2}})$ with $a_1=(1,0)a$, $a_2=(-1/2,\sqrt{3}/2)a$, $a_3=(-1/2,-\sqrt{3}/2)a$. The energy spectrum is plotted in Fig.~\ref{fig:figS6}(b). Next we consider the noncollinear magnetic order $h_{afm}=M_s(M_x s_x+M_y s_y)$ [Fig.~\ref{fig:figS6}(a)], with 
\begin{equation}
	M_x=\begin{pmatrix}
		-\frac{\sqrt{3}}{2}& 0 & 0\\
		0 & \frac{\sqrt{3}}{2} & 0\\
        0 & 0 & 0
	\end{pmatrix},
   M_y=\begin{pmatrix}
		-\frac{1}{2}& 0 & 0\\
		0 & -\frac{1}{2} & 0\\
        0 & 0 & 1
	\end{pmatrix}.
\end{equation}
Here $M_s$ denotes the strength of noncollinear magnetic order. Considering the spin degree of freedom, the model Hamiltonian has six band. Here we consider the low-energy state near $\Gamma$ pocket in Fig.~\ref{fig:figS6}(b). To further facilitate analysis, we can make the standard perturbation theory analysis focussing on the wavefunction of the lowest band near $\Gamma$ pocket 
\begin{equation}
|\psi\rangle\approx [1+\frac{i\sqrt{3}}{4}k_y-\frac{k_x^2+k_y^2}{16}+(\frac{i}{4}-\frac{\sqrt{3}}{12}k_y)k_x,1+\frac{i\sqrt{3}}{4}k_y-\frac{k_x^2+k_y^2}{16}+(-\frac{i}{4}+\frac{\sqrt{3}}{12}k_y)k_x,1]^T.
\end{equation}
Then we can obtain the momentum-space $h_{afm}(\bm{k})$ as 
\begin{equation}
h_{afm}(\bm{k})=\langle \psi|h_{afm}|\psi \rangle/\langle \psi|\psi \rangle \approx \frac{1}{48}M_s[ 2k_x k_y s_x +(k_x^2-k_y^2)s_y]
\end{equation}
One can find that the noncollinear magnetic order in real space gives rise to the momentum dependent magnetic order. Similar results are also obtained in Ref.~\cite{lee2024fermi}. To study the NSM in the superconducting state, we adopt the low-energy effective Hamiltonian as
\begin{equation}
\label{eq:model_kagome}
H_{\bm{k}}=[ta^2 (k_x^2+k_y^2)-\mu]s_0+2m a^2 k_x k_y s_x +m a^2(k_x^2-k_y^2)s_y,
\end{equation}
where $t$ is the hopping term, $m$ denotes the noncollinear magnetic order, and $s_x$, $s_y$ are spin Pauli matrices. The Hamiltonian possesses $D_{3d}$ symmetry, with $C_{3z}=(s_0-i\sqrt{3}s_z)/2$, $C_{2y}=is_y$ and $I=s_0$ (inversion) symmetries. It is straightforward to obtain the eigenenergy as $E_{\pm}=ta^2\bm{k}^2-\mu\pm ma^2\bm{k}^2$ and the corresponding spin expectation value as
\begin{equation}
(\langle s_x \rangle, \langle s_y \rangle)=(\pm \sin 2\theta, \pm \cos 2\theta),
\end{equation}
where $\theta$ is the polar angle. Based on the symmetry analysis in Table.~\ref{tab:tensor}, the nonzero spin susceptibilities are $\chi_{xx}^y=\chi_{xy}^x=\chi_{yx}^x=-\chi_{yy}^y$. Without loss of generality, we also adopt the spin-singlet pairing potential as $\hat{\Delta}=\Delta_0 is_y$. The superconducting state can be intrinsic or proximity induced. In our calculations, we also adopt the BCS temperature dependence of $\Delta_0$ with $\Delta(T)=\Delta_0\tanh(1.74\sqrt{T_c/T-1})$ and $\Delta_0=1.76k_B T_c$. 

The results are shown in Fig.~\ref{fig:figS6} with the model parameters $t=1,m=0.2,\mu=0.1$ and $\Delta_0=0.05$ and $T=0.3T_c$. In Fig.~\ref{fig:figS6}(c) we plot the BdG spectrum. Importantly, we plot Fig.~\ref{fig:figS6}(d) that displays $M_y$ from a numerical evaluation of Eq.~\eqref{eq:magnetization} as well as $\chi_{xx}^y q_x^2$ evaluated from the NSM susceptibility tensor $\chi_{ab}^c$ [Eq.~\eqref{eq:nonlinear_sus_supple}]. These show that an in-plane NSM manifests for this model. 
\end{document}